\documentclass{llncs}
\usepackage{hyperref}
\usepackage{graphicx}

\begin{document}
\title{Unstable markup:}


\subtitle{A template-based information extraction from web sites with unstable markup}


\author{Maxim Kolchin \and Fedor Kozlov}


\institute{ITMO University\\
\email{kolchinmax@niuitmo.ru}, \email{kozlovfedor@gmail.com}}



\maketitle

\begin{abstract}
This paper presents results of a work on crawling CEUR Workshop proceedings\footnote{CEUR Workshop proceedings web site, URL: \url{http://ceur-ws.org}} web site to a Linked Open Data (LOD) dataset in the framework of ESWC 2014 Semantic Publishing Challenge 2014\footnote{ESWC 2014 Semantic Publishing Challenge, URL: \url{http://2014.eswc-conferences.org/semantic-publishing-challenge}}. Our approach is based on using an extensible template-dependent crawler and DBpedia for linking extracted entities, such as the names of universities and countries.

\end{abstract}

\section{Introduction}
The work that is presented in this paper aims to provide a solution for Task 1 of ESWC 2014 Semantic Publishing Challenge\footnotemark[2]. The task is to crawl CEUR Workhop proceedings web site\footnotemark[1] and create a LOD dataset containing detailed information about workshops, proceedings volumes, papers and their authors and etc.

The source code and instructions to run the crawler are located at our Github repository\footnote{The source code and instructions, URL: \url{https://github.com/ailabitmo/sempubchallenge2014-task1}}.

\subsection{Challenges}
At first glance, the task looks a pretty straightforward, but there are several challenges that need to be solved:

\begin{itemize}
    \item the web site has a quite unstable and in some cases invalid HTML markup because of absence of a standardised and strict template for creation of pages for proceedings volumes, so it makes it harder to crawl such pages, because usually crawlers are written for web sites with fixed markup;
    \item only a small percentage of proceedings volumes uses RDFa markup and microformats are used only for volumes starting from 559th one, so at the time of writing around 49\% of volume pages don't have any metadata that could help in crawling;
    \item according to the rules of the web site, proceedings should comply with some requirements regarding numbers of invited and regular papers, therefore there are joint proceedings of several workshops. Such workshop and proceedings should be represented in the dataset accordingly;
    \item the web site includes proceeding not only in English, but also in German. In addition it's quite common practise for authors of papers written in English to use names of their universities or companies in a native language;
\end{itemize}

\section{Our approach}
We developed an extensible template-dependent crawler that uses sets of special predefined templates for each type of entity. The main aim of this templates is to cover entire variety of entity representations in HTML format. Some of templates used for extracting papers from workshops pages are:

\begin{itemize}
    \item a template based on RDFa metadata,
    \item a template based on Microformats,
    \item and two templates specific to some similar HTML markups .
\end{itemize}

When HTML page parsing begins, the crawler consecutively runs predefined templates till one of the templates returns the valid data. Validation based on template's structure. Template's parsing process extracts data from HTML page using XPath Language and regular expressions. XPath Language is used for searching text data by elements and properties in HTML markup. Regular expressions is used for extracting entity tokens from plain text. When data is extracted template parsing process converts data into ontology instances and properties. The templates are completely independent from each other and the crawler uses a mapping of the templates to the types of contents where they are applied to such as the index page, a workshop page and a publication, which makes the crawler easily extensible. More about the extensibility in the next section.

The main advantage of our approach is a flexibility of different data representations in HTML markup with usage of the same code of the crawler and support of invalid HTML.

\subsection{Architecture} 
The parser is implemented in Python and based on Grab Spider framework\footnote{Grab framework, URL: \url{http://grablib.org/}}. This framework allows to build asynchronous site crawlers. Crawler downloads all workshop's pages and papers and then runs the parsing tasks. There is a collection of specific parsers for each entity. Each parser in collection process a part of some HTML page to build properties and entity relations.

\begin{figure}[ht]
\centering
\includegraphics[width=\textwidth]{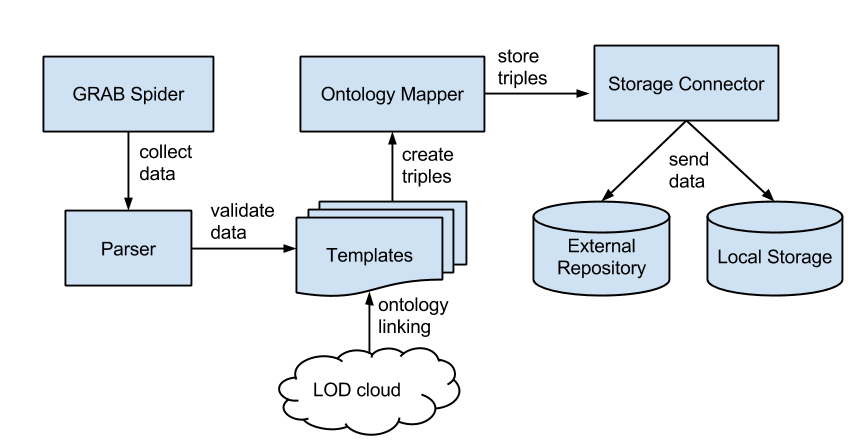}
\caption{The overall system architecture}
\label{fig:overall_sys_arch}
\end{figure}

The overall system architecture is shown in Fig. 
\ref{fig:overall_sys_arch}.  

\subsubsection{Extensibility}

The crawler provides two basic templates with the default implementations: a template for a single entity, in example a workshop page, and a template for a list of entities, in example the index page with the list of workshop names. And to add a new template, one of the basic ones should be extended by implementing the method responsible for the template matching. The rest of the work is done by the default implementations responsible for translating data to triples and writing them to disk.

\subsection{Data representation} 
To represent crawled data we use several different ontologies such as Semantic Web Conference Ontology (SWC)\footnote{Semantic Web Conference Ontology, URL: \url{http://data.semanticweb.org/ns/swc/ontology}}, Semantic Web for Research Communities ontology (SWRC)\footnote{Semantic Web for Research Communities, URL: \url{http://ontoware.org/swrc/}}, The Bibliographic Ontology (BIBO)\footnote{The Bibliographic Ontology, URL: \url{http://purl.org/ontology/bibo/}}, The Timeline Ontology (TIMELINE)\footnote{The Timeline Ontology, URL: \url{http://purl.org/NET/c4dm/timeline.owl\#}}, Friend of a Friend (FOAF)\footnote{The Friend of a Friend (FOAF), URL: \url{http://www.foaf-project.org/}}, Dublin Core (DC and DCTERMS) \footnote{Dublin Core, URL: \url{http://purl.org/dc/elements/1.1/}} and DBpedia Ontology (DBPEDIA-OWL)\footnote{DBpedia Ontology, URL: \url{http://dbpedia.org/ontology/}} and RDF Schema (RDFS)\footnote{RDF Schema, URL: \url{http://www.w3.org/2000/01/rdf-schema\#}}.

\begin{figure}[ht]
\centering
\includegraphics[width=\textwidth]{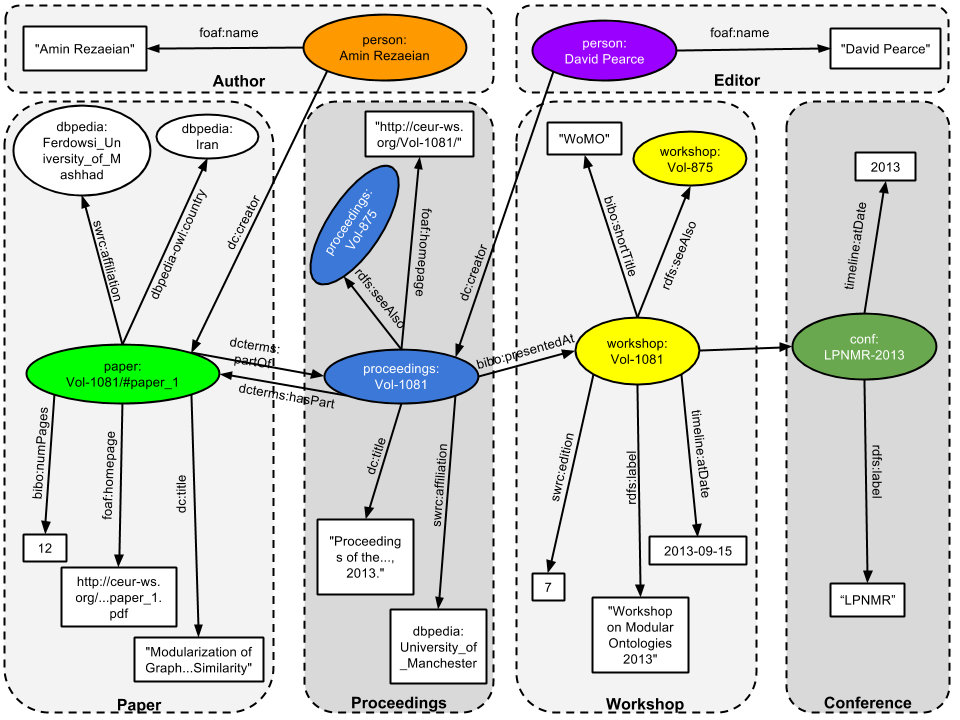}
\caption{Schema representing the crawled data}
\label{fig:schema}
\end{figure}

A part of the data representation schema is shown on Fig. \ref{fig:schema}. Representation of time and time intervals doesn't use The Event Ontology (EVENT) because it assumes inclusion of blank nodes. Since RDFLib doesn't work well with them we decided to use TIMELINE ontology instead. TIMELINE ontology provides timeline:atDate property for setting a date to an instance and \textit{timeline:beginsAtDateTime} and \textit{timeline:endsAtDateTime} properties for a time interval.

On CEUR Workshop proceedings web site some proceedings volumes has links to each other. These links usually relate a proceedings of a workshop to the previous its editions and we uses \textit{rdfs:seeAlso} property to represent this relationships.

\subsection{Specific solutions}
In most cases all problems are solved by an appropriate template, but there are some problems requiring specific solutions.

\subsubsection{Extraction of countries and affiliations}
Identification of countries and affiliations in papers was done with external datasets. In case of extracting countries parser extracts the first page from the PDF document. Country-candidates are extracted using regular expressions with predefined templates. Parser sends request with a list of country-candidates to SPARQL-endpoint of DBpedia\cite{dbpedia-swj} resource to get list of unique country's IRIs. 

The country extraction query must support different naming conventions of country-candidates. Hence the following SPARQL-query is suggested.

\begin{verbatim}
SELECT DISTINCT ?country {
    VALUES ?search { "The Netherlands" }
    ?country a dbpedia-owl:Country .
    { ?name_uri dbpedia-owl:wikiPageRedirects ?country ;
                rdfs:label ?label .
    }
    UNION
    { ?country rdfs:label ?label }
    FILTER( STR(?label) = ?search )
}
\end{verbatim}
        
Creation of properties and relations for the current paper entity is based on received list.

\subsubsection{Identification of related workshops}
As mentioned above, in most cases \textit{skos:related} property is used to relate to a previous edition of the corresponding workshops. But sometimes it's not correct. Especially in case of joint proceedings. To identify correct links we implemented the algorithm measuring similarity of two workshops based on its full name and acronym. In case of absence of an acronym we generate one from the full name's upper case characters. For example for ``Concept Extraction Challenge at Making Sense of Microposts 2013`` workshop the ``CECMSM`` acronym is generated. String similarity measurement uses the basic Ratcliff-Obershelp algorithm\cite{ratcliff1988pattern}. This algorithm was selected because it is being provided by the Python Standard Library.

\section{Conclusion}
Task 1 of Semantic Publishing Challenge 2014 is solved with developed parser based on Grab Spider framework. This parser uses SWC, SWRC, BIBO, TIMELINE ontologies, DBpedia datasets and the basic Ratcliff-Obershelp algorithm for string similarity measurement. Our approach based on templates of web site blocks, the schema representing extracted information and solutions for some specific problems.
The main advantages of our approach are flexible representation of different data templates in HTML markup and support of invalid HTML.

\subsection{Unsolved issues}
In most cases our solution works well, but there are several ``places`` where it doesn't work well and therefore may not pass some tests completely:
\begin{itemize}
\item extraction of country and university candidates from papers works only for texts consisting only of US-ASCII characters because PDFMiner\footnote{PDFMiiner, URL: \url{http://www.unixuser.org/~euske/python/pdfminer/}} which we use to extract text from PDF files doesn't work well with Unicode symbols;
\item{papers written in PostScript or HTML are completely ignored;}
\end{itemize}

\subsection{Future work}
This work can be further extended by solving the known issues and implementing additional functionality in the crawler for deeper information extraction to make the dataset more useful for further analysis. To achieve it, the following particular tasks could be done:

\begin{itemize}
\item use external repositories such as DBLP\footnote{DBLP, URL: \url{http://www.informatik.uni-trier.de/~ley/db/}}, Semantic Web Dog Food\footnote{Semantic Web Dog Food, URL: \url{http://data.semanticweb.org/}} and other open datasets to extract, link and enrich the information about authors and editors,
\item optimise extraction of authors' affiliations from papers and particular connections between the authors and the affiliations,
\item extraction of authors' e-mail addresses from papers to improve aligning of ontology instances representing authors and editors.
\end{itemize}

\subsubsection*{Acknowledgments.} This work has been partially financially supported by the Government of Russian Federation, Grant \#074-U01.

\bibliographystyle{splncs03}
\bibliography{references}

\end{document}